\documentclass[prl,twocolumn,amsmath,amssymb,superscriptaddress,showpacs]{revtex4}
\usepackage{graphicx}
\usepackage{dcolumn}
\usepackage{bm}

\usepackage{enumerate}
\newcommand{\mi}{\mathrm{i}}
\newcommand{\me}{\mathrm{e}}
\newcommand{\diff}{\mbox{d}}

\begin{document}

\title{Super-intense highly anisotropic optical transitions 
in anisotropic quantum dots}
\author{Siranush Avetisyan}
\affiliation{Department of Physics and Astronomy,
University of Manitoba, Winnipeg, Canada R3T 2N2}
\author{Pekka Pietil\"ainen}
\affiliation{Department of Physics/Theoretical Physics,
University of Oulu, Oulu FIN-90014, Finland}
\author{Tapash Chakraborty$^\ddag$}
\affiliation{Department of Physics and Astronomy,
University of Manitoba, Winnipeg, Canada R3T 2N2}

\begin{abstract}
Coulomb interaction among electrons is found to have profound effects on the 
electronic properties of anisotropic quantum dots in a perpendicular 
external magnetic field, and in the presence of the Rashba spin-orbit 
interaction. This is more evident in optical transitions, which we
find in this system to be highly anisotropic and super-intense, in particular,
for large values of the anisotropy parameter. 
\end{abstract}
\pacs{73.21.La,78.67.Hc}
\maketitle

For more than two decades, theoretical studies of quantum dots (QDs) in an 
external magnetic field \cite{maksym}, have largely focused on the properties
of dots with circular symmetry \cite{qdbook,heitmann}. Extensive investigations 
of transport and optical spectroscopy of these semiconductor nanostructures (the
{\it artificial atoms}) have revealed several important atomic-like properties
\cite{qdbook,heitmann}. In contrast, not enough is known about the electronic 
properties of anisotropic quantum dots \cite{madhav,elliptic_expt}. Another 
important direction of the QD research that is gaining popularity in recent 
years has been the role of Rashba spin-orbit interaction (SOI) 
\cite{rashba,halperin,note_Dr} in quantum dots. The importance of this 
interaction in semiconductor spintronics has been well documented in the 
literature \cite{spintro,nitta,SOI_dot}. Detailed theoretical studies of the 
influence of Rashba SOI on the electronic properties of QDs with isotropic 
confinement have already been reported earlier \cite{rashba_tc}, where 
the SO coupling was found to manifest itself mainly in multiple level crossings 
and level repulsions in the energy spectra. These were attributed to an interplay 
between the Zeeman effect and the SOI present in the system Hamiltonian. Those 
effects, in particular the level repulsions, were weak and as a result, 
would require extraordinary efforts to detect the strength of SO coupling 
\cite{tunneling} in those systems. On the other hand, by introducing anisotropy 
in a QD, we have previously shown that a major enhancement of the Rashba SO 
coupling effects can be achieved in the Fock-Darwin spectra \cite{siranush}. 
Although various approximate schemes exist to study the effects of anisotropy
on the far-infrared absorption \cite{vidar}, the role of SO coupling on the
far-infrared response \cite{serra}, or other physical properties of elliptical 
dots \cite{others}, an accurate and coherent theoretical treatment of all these 
issues, in particular, the role of Coulomb interaction, in conjunction with all 
these properties, is seriously lacking. Here we demonstrate that in the presence 
of the Coulomb interaction among the electrons, and combined with the Rashba SOI, 
the eccentricity of the QD is responsible for major modification of the electron 
energy spectra, which clearly manifests itself in super-intense, and highly 
anisotropic optical transitions that is vastly different from those that are
commonly observed in an isotropic QD. 

\begin{center}
\begin{figure}
\includegraphics[width=7cm]{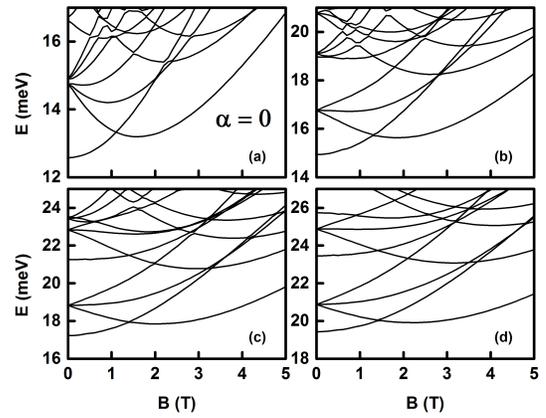}
\caption{\label{fig:alpha=0} Magnetic field dependence of the low-lying
two-electron energy levels of an elliptical dot without the Rashba SOI
$(\alpha=0)$. The results are for $\omega^{}_x=4$ meV and (a)
$\omega^{}_y=4.1$ meV, (b) $\omega^{}_y=6$ meV, (c) $\omega^{}_y=8$ meV,
and (d) $\omega^{}_y=10$ meV.
}
\end{figure}
\end{center}

Until now, interacting electrons in elliptical QDs have been studied by means of
perturbative approaches \cite{elliptic_int}. In what follows, we present a 
non-perturbative, exact diagonalization scheme to treat interacting electrons in
anisotropic quantum dots. Our complete single-particle Hamiltonian of an electron 
moving in the $xy$-plane and subjected to an external perpendicular magnetic field 
with the vector potential $\mathbf A=\frac12B(-y,x)$ is
\begin{eqnarray*}
{\cal H}&=&\frac1{2m^{}_e}\left(\mathbf p-\frac ec\mathbf A\right)^2
+\tfrac 12 m^{}_e\left(\omega_x^2 x^2+\omega_y^2 y^2\right)
\\ &&
+\frac\alpha\hbar\left[\bm\sigma\times
\left(\mathbf p-\frac ec\mathbf A\right)\right]^{}_z
+\tfrac12 g\mu^{}_B B\sigma^{}_z.
\end{eqnarray*}
The first two terms on the right hand side describe a two-dimensional harmonic 
oscillator confined by an elliptic potential \cite{madhav}. The next term takes care 
of the SOI while the last one is for the Zeeman coupling. 
In order to treat the Coulomb interaction we rearrange the terms in the Hamiltonian
into three parts
\begin{eqnarray*}
{\cal H}^{}_\Lambda&=&\frac1{2m^{}_e}
\left(p_x^2+p_y^2+\Omega_x^2x^2+\Omega_y^2y^2\right); \ \
{\cal H}^{}_Z=\tfrac12 g\mu^{}_B B\sigma^{}_z \\
{\cal H}^{}_R&=&\tfrac12\omega^{}_c(yp^{}_x-xp^{}_y)
	+\frac\alpha\hbar\left[\sigma^{}_x\left(p^{}_y-\frac{eB^{}_\perp}{2c}\,x\right) 
\right. \\
&&\left. -\sigma^{}_y\left(p^{}_x+\frac{eB^{}_\perp}{2c}\,y\right)\right],
\end{eqnarray*}
where ${\cal H}^{}_\Lambda$ describes a two-dimensional spinless harmonic 
oscillator, the Zeeman coupling ${\cal H}^{}_Z$ introduces the spin and ${\cal 
H}^{}_R$ deforms the simple Cartesian phase space of the operators ${\cal H}^{}_\Lambda$ 
and ${\cal H}^{}_Z$. We have also introduced the cylotron frequency $\omega^{}_c
=eB/{m^{}_ec} $ and the oscillator frequences $\Omega_{x,y}^2=m_e^2\left(\omega_{x,y}^2
+\tfrac14 \omega_c^2\right).$ The eigenstates $|\lambda\rangle$ of the oscillator 
Hamiltonian ${\cal H}^{}_\Lambda$ are just direct products $|n^\lambda_x\rangle
|n^\lambda_y\rangle$ of the two harmonic oscillator states represented by the quantum 
numbers $n^\lambda_{x,y}$. Inclusion of the Zeeman term is also straightforward: we 
multiply the states $|\lambda\rangle$ with the eigenstates $|s^{}_z\rangle$ of the Pauli 
spin matrix $\sigma^{}_z$ yielding the states $|\xi\rangle=|\lambda^\xi\rangle|s^\xi_z
\rangle.$ Finally, the effects of the operator ${\cal H}^{}_R$  are incorporated by 
diagonalizing it in the base spanned by the eigenstates $|\xi\rangle$ of the combination 
${\cal H}^{}_\Lambda+{\cal H}^{}_Z$. Thus the eigenstates $|\gamma\rangle$ of the total 
single electron Hamiltonian ${\cal H}$ are experessed as superpositions of the states 
$|\xi\rangle$.

\begin{center}
\begin{figure}
\includegraphics[width=7cm]{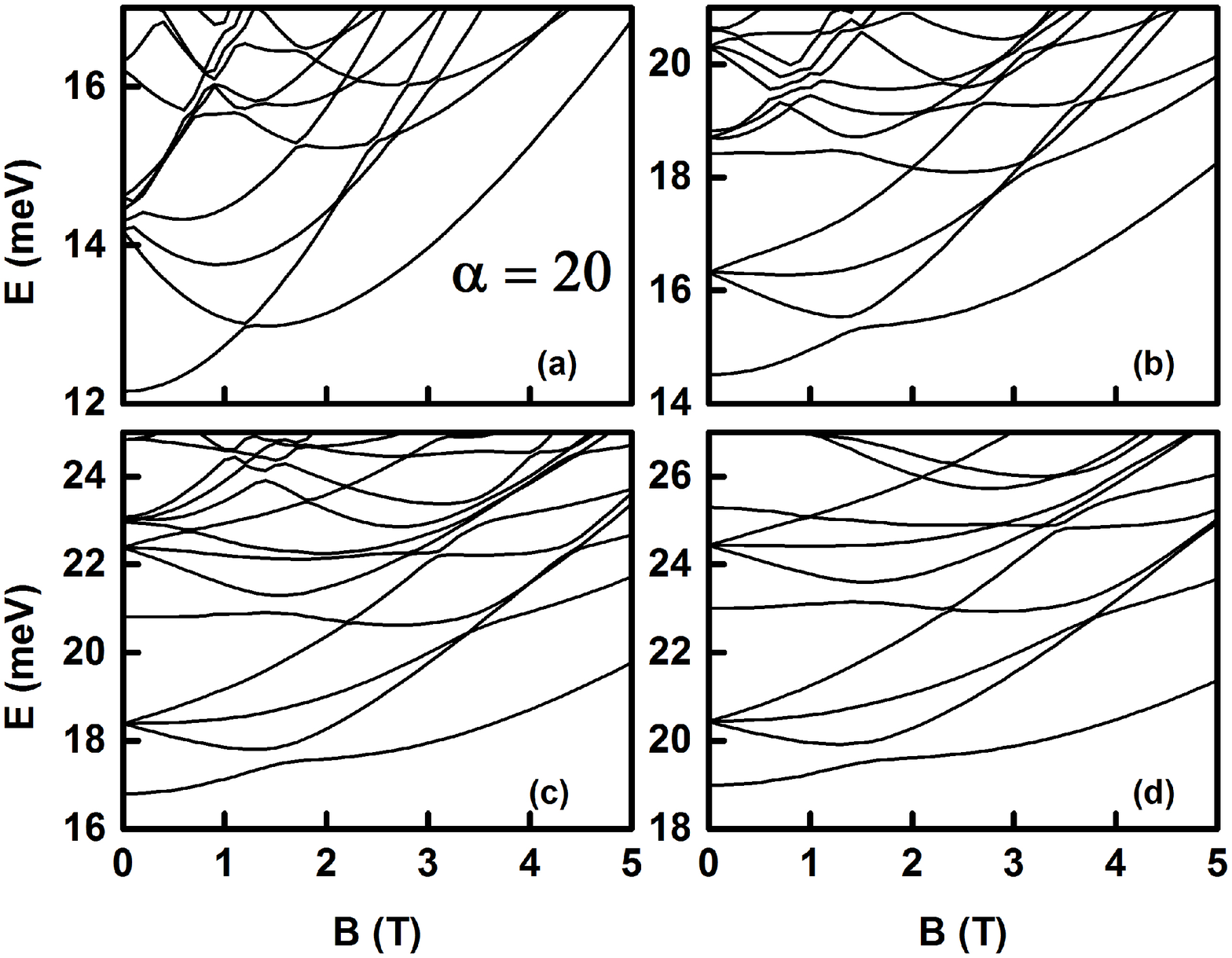}
\caption{\label{fig:alpha=20} Same as in Fig.~\ref{fig:alpha=0}, but for
$\alpha=20$ meV.nm.}
\end{figure}
\end{center}

To handle the mutual interactions between electrons, we work in the occupation number 
representation based on the eigenstates of the Hamiltonian ${\cal H}$. Then the main 
task is to evaluate the two-body matrix elements
\begin{eqnarray*}
\langle\gamma^{}_1\gamma^{}_2|V|\gamma^{}_3\gamma^{}_4\rangle
&=&\int\diff\mathbf x^{}_1\,\diff\mathbf x^{}_2\,
\Phi_{\gamma^{}_1}^\ast(\mathbf x^{}_1)\Phi_{\gamma^{}_2}^\ast(\mathbf x^{}_2)\\
&&\times V(|\mathbf x^{}_1-\mathbf x^{}_2|)
\Phi^{}_{\gamma^{}_3}(\mathbf x^{}_2)\Phi^{}_{\gamma^{}_4}(\mathbf x^{}_1),
\end{eqnarray*}
where the wave functions $\Phi^{}_\gamma(\mathbf x)$ correspond to the eigenstates 
$|\gamma\rangle$ of ${\cal H}$ and the integrals over the variables $\mathbf x$ 
include also summation over the spin degrees of freedom. The expansion of the 
functions $\Phi^{}_\gamma$ in terms of the wave functions corresponding to the 
eigenstates $|\xi\rangle$ of ${\cal H}^{}_\Lambda+{\cal H}^{}_Z$ leads to evaluations 
of two-body matrix elements between the states $|\xi\rangle$. Since the electrons 
act via the Coulomb potential $V(|\mathbf x|)=V^{}_C(r)=e^2/\epsilon r$, where
$\epsilon$ is the background dielectric constant, the summations over spin degrees 
of freedom yield only Kronecker delta's of the $s^{}_z$ quantum numbers and we are 
left with the matrix elements
\begin{eqnarray*}
\langle\lambda^{}_1\lambda^{}_2|V|\lambda^{}_3\lambda^{}_4\rangle
&=&\int\diff\mathbf r^{}_1\,\diff\mathbf r^{}_2\,
\psi_{\lambda^{}_1}^\ast(\mathbf r^{}_1)\psi_{\lambda^{}_2}^\ast(\mathbf r^{}_2)\\
&&\times V(|\mathbf r^{}_1-\mathbf r^{}_2|)
\psi^{}_{\lambda^{}_3}(\mathbf r^{}_2)\psi^{}_{\lambda^{}_4}(\mathbf r^{}_1)
\end{eqnarray*}
between pairs of the single-particle oscillator wave functions. In isotropic parabolic 
dots with mutual Coulomb interactions we could use the explicit algebraic formula 
\cite{rashba_tc}, but in elliptical confinements we have to resort to numerical
computations. Perhaps the most cost-effective way is to do the evaluation via the 
Fourier transforms
\begin{eqnarray*}
\tilde\Psi^{}_{\mu\nu}(\mathbf k)
&=&\int\diff\mathbf r\,\me^{\mi\mathbf k\cdot\mathbf r}
\psi_{\mu}^\ast(\mathbf r)\psi^{}_{\nu}(\mathbf r) \\
\tilde V(\mathbf k)
&=&\int\diff\mathbf r\,\me^{\mi\mathbf k\cdot\mathbf r}V(\mathbf r)
\end{eqnarray*}
of the products of the wave functions and the interaction. A straightforward algebra 
yields the expression
$$\langle\lambda^{}_1\lambda^{}_2|V|\lambda^{}_3\lambda^{}_4\rangle
=\frac1{(2\pi)^2}\int\diff\mathbf k\,
\tilde\Psi_{\lambda^{}_4\lambda_1}^\ast(\mathbf k)
\tilde\Psi^{}_{\lambda^{}_2\lambda^{}_3}(\mathbf k)
\tilde V(\mathbf k).$$
Numerical computation of this final two-fold integral is a relatively fast operation.

\begin{center}
\begin{figure}
\includegraphics[width=7cm]{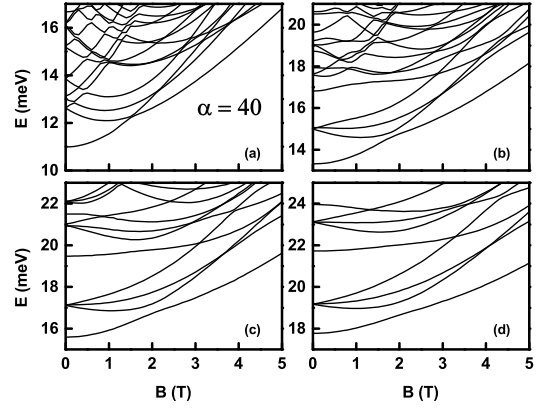}
\caption{\label{fig:alpha=40} Same as in Fig.~\ref{fig:alpha=0}, but for
$\alpha=40$ meV.nm.}
\end{figure}
\end{center}

Since for the Coulomb interactions we know the Fourier transform to be 
$\tilde V^{}_C(\mathbf k)=2\pi e^2/\epsilon\,k,$ 
we are left with the evaluation of the Fourier transforms $\tilde\Psi^{}_{\mu\nu}
(\mathbf k)$. We have experimented with two practically equally efficient methods: the 
first one is fully generic and applicable to any system while the second one is 
restricted to elliptical confinements. The generic method is based on the observation 
that the Fourier transform $\tilde\Psi^{}_{\mu\nu}(\mathbf k)$ can in fact be written 
as the matrix element of the exponential of the position operator $\mathbf
r^{}_{\mathrm{op}}$, $\tilde\Psi^{}_{\mu\nu}(\mathbf k)= \left\langle \mu\left|
\me^{\mi\mathbf k\cdot\mathbf r^{}_{\mathrm{op}}}\right|\nu\right\rangle$. Since the 
components $x^{}_\mathrm{op}$ and $y^{}_\mathrm{op}$ of the position operator commute 
we actually need the matrix elements of the exponential operators $\exp\left(\mi 
x^{}_\mathrm{op}\right)$ and $\exp\left(\mi y^{}_\mathrm{op}\right)$. These in turn 
are easily evaluated by diagonalizing the matrix $X$ with matrix elements 
$X^{}_{\mu\nu}=\left\langle \mu\left| x^{}_{\mathrm{op}} \right|
\nu\right\rangle$ and applying the inverse unitary transformation taking $X$ to the 
diagonal form to the exponentiated diagonal, together with the similar 
procedure for the operator $y^{}_\mathrm{op}$.

\begin{center}
\begin{figure}
\includegraphics[width=7cm]{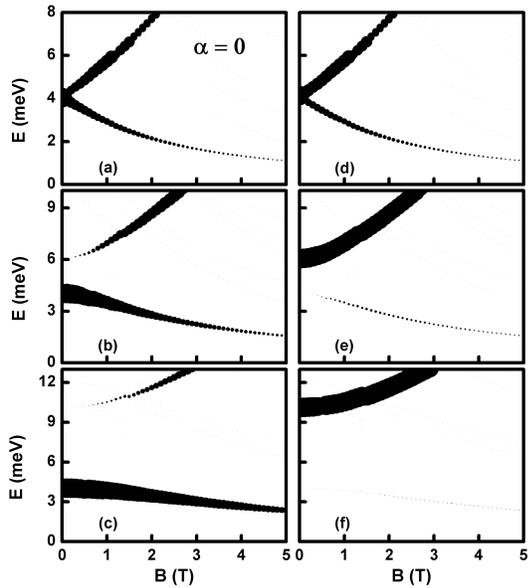}
\caption{\label{fig:trans1} Optical absorption (dipole allowed) specta
of elliptical QDs with $\alpha=0$ meV.nm for various choice of anisotropy
parameters: (a) $\omega^{}_x=4$ meV, $\omega^{}_y=4.1$, (b)
$\omega^{}_x=4$ meV, $\omega^{}_y=6$ meV, and (c) $\omega^{}_x=4$,
$\omega^{}_y=10$. The polarization of the incident radiation is along the
$x$-axis. The parameters for (d)-(f) are the same, except that the incident
radiation is polarized along the $y$-axis. The areas of the filled circles are
proportional to the calculated absorption cross-section.
}
\end{figure}
\end{center}

In the second approach we take advantage of the fact that the single-particle wave 
functions are products of two Hermite functions, one with the $x$-coordinate and 
the other with the $y$-coordinate as the argument. This implies that the product 
$\psi_{\mu}^\ast(\mathbf r) \psi^{}_{\nu}(\mathbf r)$ to be transformed factorizes 
to a product of two functions depending on $x$ and $y$, respectively. In fact we 
can do the resulting one-dimensional transforms yielding
$$\tilde\Psi^{}_{\mu\nu}(\mathbf k )
=\tilde{\cal G}^x_{n^\mu_x,n^\nu_x}(k^{}_x)
\tilde{\cal G}^y_{n^\mu_y,n^\nu_y}(k_y),$$
where the functions $\tilde{\cal G}^{x,y}_{i,j}(k)$ are given by
\begin{eqnarray*}
\tilde{\cal G}^{x,y}_{ij}(k)
&=&(-1)^m\sqrt{\frac{n!}{(n+\xi)!}}\,(\beta^{}_{x,y} k)^\xi\,
\me^{-\frac12(\beta^{}_{x,y} k)^2} \\
&&\times L_n^\xi\left((\beta^{}_{x,y} k)^2\right) (\delta^{}_{p,0}+\mi\delta^{}_{p,1}).
\end{eqnarray*}
In the above formulas the symbols $n^\mu_{x,y}$ and $n^\nu_{x,y}$
stand for the $x$ and $y$ oscillator quantum numbers of the
states labelled by $\mu$ and $\nu$. We have also introduced
the shorthand notations $\beta^{}_{x,y}=\sqrt{\frac\hbar{2m^{}_e\omega^{}_{x,y}}}, 
\, n=\min(i,j),\, \xi=|i-j|,\, p=\xi\,\mbox{mod\,}2.$
Although both approaches introduced here have their merits, it should be noted
that the first method is somewhat more general and applicable to any systems, while
the second method works only for the harmonic oscillator basis. The second method is 
however computationally slightly faster than the first.

\begin{center}
\begin{figure}
\includegraphics[width=7cm]{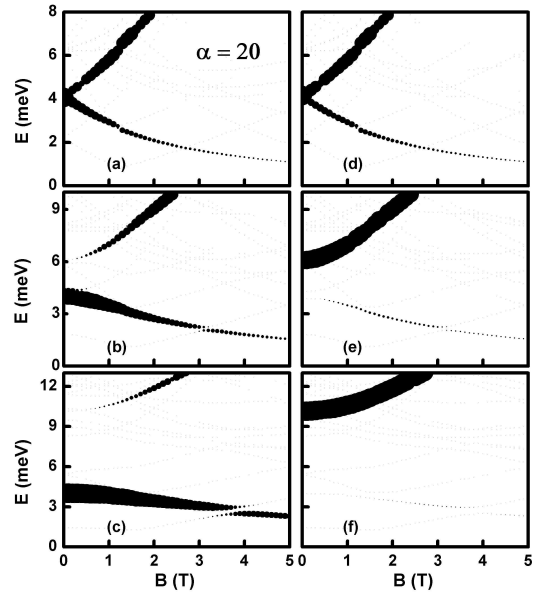}
\caption{\label{fig:trans2} Same as in Fig.~\ref{fig:trans1}, but for
$\alpha=20$ meV.nm.}
\end{figure}
\end{center}

In our numerical studies that follow, we have used the parameters corresponding to
InAs QD \cite{siranush}, where strong SOI was reported experimentally \cite{SOI_dot}.
The results for the energy spectra are displayed in Figs.~(\ref{fig:alpha=0} -
\ref{fig:alpha=40}) for various values of the SO coupling strength $\alpha$ and
the anisotropy. In the absence of the external magnetic field and the SOI, neither the 
total spin $S$ nor its $z$-component $S_z$ appear in the {\it full} Hamiltonian (with 
Coulomb interactions). We therefore expect the two-electron systems to consist of $S=0$ 
spin singlets and $S=1$ spin triplets. The energy spectra of Fig.~1 indeed confirm that
to be the case: the dispersions form bunches of one and three lines, the latter of 
which diverges due to the Zeeman splitting when the magnetic field increases.
Perhaps the most noticeable feature shown in Fig.~1 is the singlet-triplet transition 
of the ground state for magnetic fields slightly above 1 Tesla. The origin
of this crossing of the dispersion lines can be traced to the crossing of the second 
and third lowest energy levels of the single-electron systems \cite{siranush}.

Just as for the circular QD, the spin singlet-triplet transition (at B$\simeq 
1.5$ Tesla in Fig.~\ref{fig:alpha=0}) is the only transition in the ground state. 
The critical field where the transition takes place is somewhat dependent 
on the method of calculation and the choice of material parameters \cite{elliptic_int}. 
The main role of the Coulomb interaction is the upward shift of the spectral lines 
and lifting of the accidental degeneracies. Surprisingly, the interaction also practically
freezes the movement of the singlet-triplet transition point to higher fields when the 
eccentricity increases. In the absence of the electron-electron interaction the transition 
point shifts about two Teslas whereas in the presence of the interactions the shift
is only few tenths of a Tesla with the same eccentricities.

\begin{center}
\begin{figure}
\includegraphics[width=7cm]{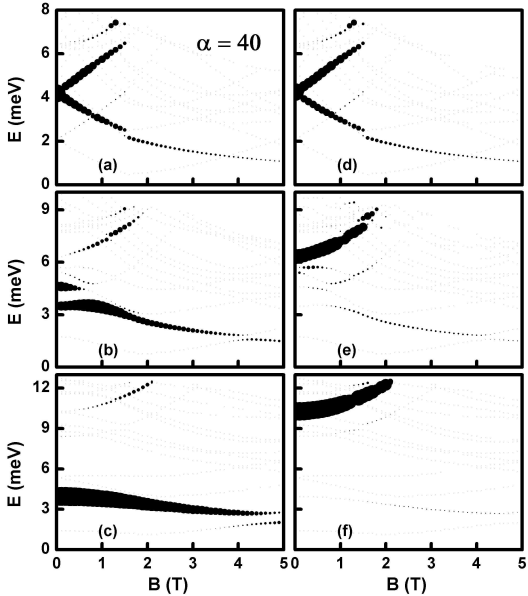}
\caption{\label{fig:trans3} Same as in Fig.~\ref{fig:trans2}, but for
$\alpha=40$ meV.nm.}
\end{figure}
\end{center}

When the SOI is turned on (Fig.~2 and Fig.~3) most of the characteristic features of 
Fig.~1 survive. However, since the SOI can mix spin-up and spin-down single-particle 
states, neither $S$ nor $S_z$ are any more good quantum numbers. This is clearly 
evident in the singlet-triplet transition which transforms to anticrossing in the 
presence of the SOI. Several similar kind of crossing-anticrossing conversions
can also be seen higher in the spectra.

In Figs.~4--6 we show the the absorption cross sections for the dipole allowed 
transtions from the ground states corresponding to the energy spectra of Figs.~1--3, 
(a,b,d). We explore the cases where the incident radiation is polarized along the $x$ 
and $y$-direction. Since dipole absorptions involve only one electron we are effectively 
probing the single-particle properties of the dot, in particular, the oscillator 
strengths along the $x$ and $y$-directions. Consequently we expect the absorption 
sepectra to resemble approximately the spectra of the one-particle system. This indeed 
seems to be the case. Except for the case of almost isotropic QDs [(a) and (c)], the 
optical transitions are clearly highly anisotropic. For example, because the 
$y$-polarization probes for oscillations along the $y$-direction the related transitions 
go mostly to the upper mode, i.e., the favored transition energies are 4, 6 and
10 meV in Figs.~(d)-(f). The resulting transitions are therefore super-intense, unlike 
in isotropic QDs. There are also weak-intensity transitions to the lower mode. This is 
due to the magnetic field and the SOI, both of which distort the confinement ellipsoid.
There are of course some notable deviations from the single-electron case. For example, 
because the Coulomb interaction couples several non-interacting states there can be many 
more allowed transitions from a given interacting state than from a non-interacting one 
resulting in different absorption intensities. 

To summarize: we have reported here very comprehensive and accurate studies of 
anisotropic quantum dots with interacting electrons in the presence of the Rashba SOI. 
We have shown here that the Coulomb interaction in the presence of the spin-orbit 
coupling has a very strong effect, particularly in the presence of strong anisotropy. 
This is clearly seen in the optical absorption spectra which is super-intense and highly
anisotropic. The spectra derived here are entirely different from the ones observed
thus far in isotropic QDs. Our present work can be generalized, in a straightforward 
manner, to include more interacting electrons in the QD. The energy spectra and the optical 
transitions with more electrons will undoubtedly be very complex. However, the basic 
properties uncovered here will remain intact. Those studies will 
be the subject of our future publications.

The work was supported by the Canada Research Chairs Program of the
Government of Canada.

\end{document}